\documentclass[epj]{svjour}
\newcommand{\npb}[3]{Nucl.~Phys.~B #1 (#2) #3}
\newcommand{\plb}[3]{Phys.~Lett.~B #1 (#2) #3}
\newcommand{\prd}[3]{Phys.~Rev.~D #1 (#2) #3}

\newcommand{\zpc}[3]{Z.~Phys.~C #1 (#2) #3}
\newcommand{\cpc}[3]{Comput.~Phys.~Commun. #1 (#2) #3}
\newlength{\caheight}
\multiply\caheight by 7
\newlength{\secondpar}
\divide\secondpar by 3
\newlength{\firstpar}
\multiply\firstpar by 2

\usepackage[dvips]{graphicx}
\usepackage{amssymb}
\usepackage{xspace}
\begin{document}
%
\title{Radiative four--fermion processes at LEP2}
\author{G.~Montagna\inst{1}, M.~Moretti\inst{2}, 
O.~Nicrosini\inst{3}, M.~Osmo\inst{1}, F.~Piccinini\inst{3}
}                     
%
%
\institute{Dipartimento di Fisica Nucleare e Teorica - Universit\`a
di Pavia 
and INFN - Sezione di Pavia, via A. Bassi 6, Pavia, Italy
 \and 
Dipartimento di Fisica - Universit\`a di Ferrara 
                  and INFN - Sezione di Ferrara, Ferrara, Italy
 \and
INFN - Sezione di Pavia, via A. Bassi 6, Pavia, Italy 
and Dipartimento di Fisica Nucleare e Teorica - Universit\`a
di Pavia
}
\date{Received: date / Revised version: date}
%
\abstract{
The production of four fermions plus a visible photon in electron-positron 
collisions is analyzed, with particular emphasis on the LEP2 energy range. 
The study is based on the calculation of exact 
matrix elements, including the effect of fermion masses. 
In the light of the present measurements performed at LEP, 
triple and quartic anomalous gauge couplings are taken 
into account. Due to the presence of a visible photon in the final state, 
particular attention is paid to the treatment 
of higher-order QED corrections. Explicit results for integrated cross 
sections and differential distributions are shown and commented.
The features of the Monte Carlo program {\tt WRAP}, used to perform the
calculation and available for experimental analysis, are described.  
\PACS{{12.15.Ji}{} \and
      {13.40.Ks}{} \and
      {13.85.Hd}{}
     } 
} 
\maketitle
\section{Introduction}
\label{intro}
One of the main goals of electroweak physics at LEP2 is the study of the 
properties of the $W$ bosons~\cite{yr96}.
The center of mass (c.m.) energy above the threshold 
of $W$-pair production offers the possibility to extract
information about the mass and couplings 
of the $W$ boson from the analysis
of four-fermion ($4f$) final states. As well known, because of the precision
of the experimental measurements,
radiative corrections to $e^+ e^- \to 4f$ processes 
are needed in order to provide suitable
theoretical predictions~\cite{yr96,lep2mc}. 
From this point of view, events with 
four fermions plus a visible photon, i.e 
$e^+ e^- \to 4f + \gamma$,  are a building block of the 
full $O(\alpha)$ electroweak corrections to $4f$ processes, providing
the hard bremsstrahlung contribution.
Furthermore, radiative $4f$ processes are also an interesting physics  
subject by themselves, since
the luminosity achieved at LEP makes them 
directly accessible to the experimental investigation, as recently discussed 
in refs.~\cite{l3quad,opalqagc,ss}, where first results on the 
measurement of the $W^+W^-\gamma$ cross section have been reported.

A very peculiar feature of the processes under consideration is that they give the 
opportunity of directly testing 
the non-abelian structure of the gauge boson
self-interactions. Actually, as other 
processes studied at LEP, such 
as $e^+ e^- \to W^+ W^- \to 4f$, $e^+ e^- \to W e \nu_e$ 
(single-$W$ production) and $e^+ e^- \to \nu\bar{\nu}\gamma$,  
$e^+ e^- \to 4f + \gamma$ reactions are sensitive to trilinear gauge couplings 
(TGC). More importantly, they can be used to test quartic gauge couplings 
(QGC), since they are, together with $\nu {\bar \nu} \gamma\gamma$
final states, the only accessible LEP2 processes
that contain quartic gauge boson 
vertices with at least one photon at tree-level~\cite{ss}. 
Quartic vertices involving only massive gauge bosons 
give rise to six-fermion final states and 
are outside the sensitivity of LEP, 
being accessible only at the energies of a 
future $e^+ e^-$ Linear Collider (LC). Both charged current (CC) 
radiative $4f$ processes, mediated by two $W$-bosons, and
neutral current (NC) ones, mediated by two
$Z$-bosons, are in principle 
suitable to examine the effect of possibly anomalous gauge 
couplings (AGC).
In this work particular attention is paid to CC processes, because  
of the larger cross section of 
$WW\gamma$ events with respect to $ZZ\gamma$ final states.

The first tree-level calculations of $e^+ e^- \to 4f + \gamma$ processes were performed in
refs.~\cite{grc,cm}. In these papers all electroweak contributions
as well as fermion mass effects were accounted for by exploiting different 
approaches to the automatic calculation of the exact tree-level matrix
element~\cite{grace,alpha}. Since then, some event generators for 
the simulation of $4f +\gamma$ events in 
$e^+ e^-$ collisions have been developed: {\tt RacoonWW}~\cite{ddrw}, 
a generator based on the calculation of all $4f+\gamma$ 
final states in the massless approximation~\cite{ditt99}; 
{\tt CompHEP}~\cite{comphep} and {\tt grc4f}~\cite{grc4f}, which are general-purpose 
packages relying upon the automatic calculation 
of tree-level amplitudes (including fermion 
masses) and phase-space; {\tt Helac/Phegas}~\cite{helac}, a 
program implementing a recursive algorithm for 
the calculation of the scattering amplitudes. 
The interested reader is referred to ref.~\cite{lep2mc} 
for more details. In addition to the above computational tools, also a 
calculation of the massive matrix element of $e^+ e^- \to 4f + \gamma$ processes 
have recently appeared in the literature~\cite{koljeg}, accompanied 
by a detailed phenomenological analysis of fermion mass effects in 
$4f$ and $4f+\gamma$ final states at LEP2 energies~\cite{kj2}.  
In all the theoretical studies devoted insofar to 
$4f + \gamma$ production, the effects of quartic anomalous 
gauge couplings (QAGC), which are a window on the 
mechanism of spontaneous symmetry breaking~\cite{book} 
and are presently of special experimental interest, have not been 
taken into account. Actually, recent phenomenological studies 
on the subject of QAGC 
at high-energy $e^+ e^-$ colliders have been performed by considering 
three-vector boson $WW\gamma, ZZ\gamma, Z\gamma\gamma$ production and treating 
$W,Z$ particles in 
the on-shell approximation~\cite{stirlingwer,bela}. 
Anomalous quartic couplings in $\nu\bar\nu \gamma \gamma$ production 
via $WW$ fusion have been analyzed in ref.~\cite{swplb}.  
Experimental 
searches for QAGC at LEP rely upon
the theoretical results of refs.~\cite{stirlingwer,bela,swplb,bb,an} and make use 
of the computational tools of refs.~\cite{stirlingwer,swplb}.

In the light of the present situation and in view of 
future measurements at LC, a full calculation 
of $e^+ e^- \to 4f + \gamma$ processes, including the effects of AGC and 
of the most important radiative corrections, is desirable. 
This task is accomplished in the present paper, 
by presenting the new event generator 
{\tt WRAP} (W Radiative process with {\tt ALPHA}~\cite{alpha} and Pavia) 
for the simulation of $4f + \gamma$ processes at $e^+ e^-$ 
colliders. The program is based on the calculation of exact matrix elements, 
including 
the effect of fermion masses, both for
CC and NC processes. 
Charged trilinear anomalous gauge couplings 
(TAGC) and the genuinely QAGC, 
i.e. those giving no contributions to 
trilinear vertices, are included in the calculation, as well as
the large effect of initial state radiation (ISR). 
A tuned comparison between the predictions of a preliminary version 
of {\tt WRAP} and of the other two event generators {\tt RacoonWW} and {\tt Helac/Phegas}
can be found in ref.~\cite{lep2mc}.

The paper is organized as follows. In Sect.~2 the main features of the
the calculation are described. After the description of the treatment 
of the multi-particle phase-space, the theoretical details concerning 
the calculation of the exact matrix elements, the implementation of 
anomalous gauge couplings and ISR are given.
A sample of numerical results as obtained by means of the Monte Carlo 
{\tt WRAP} is presented in Sect.~3, paying particular attention to the 
contribution of fermion masses, to the 
impact of ISR and to the effects of AGC at LEP2 and LC energies. 
Conclusions and possible perspectives are drawn in Sect.~4.

\begin{figure*}[t]
\begin{center}
\includegraphics[bb= 128 656 547 725,scale=1.0]{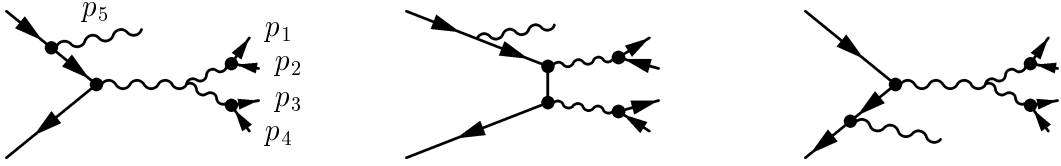}
\caption{Example of Feynman diagrams for photon radiation from the initial 
state.}
\label{fig1_epj}
\end{center}
\end{figure*}


\begin{figure*}[t]
\begin{center}
\includegraphics[bb= 116 637 532 717,scale=1.0]{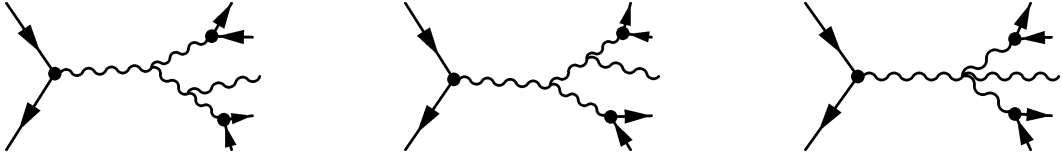}
\caption{Example of Feynman diagrams for photon radiation from $W$ bosons.}
\label{fig2_epj}
\end{center}
\end{figure*}

\begin{figure*}[t]
\begin{center}
\includegraphics[bb= 152 650 433 726,scale=1.0]{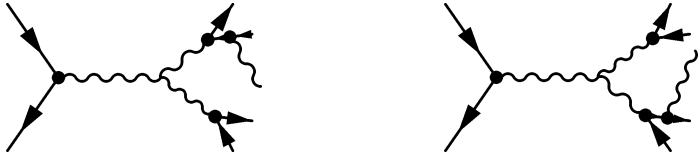}
\caption{Example of Feynman diagrams for photon radiation from final-state 
fermions.}
\label{fig3_epj}
\end{center}
\end{figure*}

\section{Features of the calculation}
\label{sec:1}
\subsection{Phase-space integration}
\label{sec:2}

The kinematics of the $2\to 5$ particles processes has been treated 
generating the $5$-body phase-space recursively, since the process can be seen
as production and subsequent decay of a pair of massive gauge bosons.

Concerning CC like processes, the configurations of interest are related 
to a 
photon emitted from the initial state (see Fig.~\ref{fig1_epj}), 
from an intermediate $W$-boson (see Fig.~\ref{fig2_epj}) and from the final
state charged fermions (see Fig.~\ref{fig3_epj}).
As far as emission from the initial state is concerned, the 
adopted phase-space decomposition reads as follows:

\begin{equation}
\begin{array}{lll}
d\Phi_5  = & (2\pi)^{6}&
d\Phi_3(P;p_5,Q_{V_1},Q_{V_2})d\Phi_2(Q_{V_1};p_1,p_2)\\
& & d\Phi_2(Q_{V_2};p_3,p_4)dQ^2_{V_1} dQ^2_{V_2},
\end{array}
\label{phsp1}
\end{equation}
where $V_1$ and $V_2$ indicate the $W$ gauge bosons, the momenta 
$p_i$ with $i = 1, \ldots, 4$ stand for the momenta of the final state 
fermions and $p_5$ is the photon momentum. 
The eleven independent variables have been chosen to be:
\begin{itemize}
\item photon variables $E_{\gamma}$, $\theta_{\gamma}$ and
$\phi_{\gamma}$ in the c.m. frame;
\item invariant mass squared $Q^2_{V_1}$ and $Q^2_{V_2}$;
\item three $\theta$ and $\phi$ angle pairs in the rest frame of each decaying
``particle'', namely in the frames given by the conditions ${\bf P} - {\bf p}_5= {\bf 0}$, 
${\bf Q}_{V_1}={\bf 0}$ and ${\bf Q}_{V_2}={\bf 0}$.
\end{itemize}

When the photon is
emitted from the final state, which means for a CC process 
from a virtual $W$ or from a virtual charged fermion, 
the following decomposition can be conveniently considered:

\begin{equation}
\begin{array}{lll}
d\Phi_5 = &(2\pi)^{6}&
d\Phi_2(P;Q_{V_1},Q_{V_2})d\Phi_3(Q_{V_1};p_5,p_1,p_2)\\
& & d\Phi_2(Q_{V_2};p_3,p_4)dQ^2_{V_1}dQ^2_{V_2}.
\end{array}
\label{phsp31}
\end{equation}

In the case of the photon emitted from an internal gauge boson the independent 
variables can be chosen as follows:
\begin{itemize}
\item photon variables $E_{\gamma}$, $\theta_{\gamma}$ and
$\phi_{\gamma}$ in the c.m. frame;
\item invariant mass squared $Q^2_{V_1}$ and $Q^2_{V_2}$;
\item one $\theta$ and $\phi$ $W$-angle pair in the c.m. frame;
\item two $\theta$ and $\phi$ angle pair for ${\bf p}_1$ and 
${\bf p}_3$ in the rest frame of the bosons $V_1$ and $V_2$, respectively.
\end{itemize}
In the case of photon emission from a final state fermion 
the following independent variables can been adopted:
\begin{itemize}
\item invariant mass squared $Q^2_{V_1}$ and $Q^2_{V_2}$;
\item one $\theta$ and $\phi$ $W$-angle pair in the c.m. frame;
\item one $\theta$ and $\phi$ angle pair for ${\bf p}_3$ in the rest frame 
of the boson $V_2$;
\item energies of $p_1$ and $p_5$ momenta in the c.m. frame;
\item azimuthal angle $\phi$ of $p_1$ in the rest frame of $V_1$;
\item $\phi_\gamma$ in the rest frame of the radiating fermion;
\item $\cos\theta_{\gamma-f}$ in the c.m. frame ,
\end{itemize}
where $\theta_{\gamma-f}$ is the relative angle 
between the radiating fermion and the photon.

An analogous phase-space parameterization has been implemented in {\tt WRAP} for the case of 
NC processes, neglecting of course the channels related to 
the photonic emission from internal lines. 
The above phase-pace decompositions, iterated for each possible radiation
pattern, give rise to several channels, depending on the final state
considered. 

In the previous formulas $d\Phi_n$ represents the element of $n$-body phase-space
given by:

\begin{equation}
d\Phi_n(P;p_1,\ldots ,p_n)\hspace{.2cm}
=\hspace{.2cm}(2\pi)^4\hspace{.1cm}\delta^4(P-\sum_{i=1}^n 
p_i)\prod_{i=1}^n\frac{d^3~p_i}{(2\pi)^3~2E_i}. 
\end{equation}

The code works taking into account all the configurations discussed
above according to a standard multi-channel Monte Carlo approach~\cite{mult}. 

In order to perform an efficient event generation, the peaking behaviour of 
the matrix element has been treated in the following way: 
\begin{itemize}
\item the squared invariant masses of the massive gauge bosons $V$
are sampled according to a
Breit-Wigner
distribution centered around $M^2_V$, while the photon propagator is sampled
according to the $1/Q^2$ distribution;
\item the infrared divergence is sampled according to a $1/E_{\gamma}$
distribution;
\item the collinear peak arising from the photon emission due to an
external 
charged fermion is taken under control by sampling it with a distribution
proportional to~$1/(1-\beta\cos\theta)$, where $\theta$ is the separation
angle 
between the radiating fermion and the photon, and 
$\beta = \sqrt{1 - m^2/E^2}$, $m$ and $E$ being the mass and the energy of 
the fermion respectively. 
\end{itemize}

For a realistic account of gauge boson properties, and to avoid 
integration singularities, it is mandatory to
include the gauge boson width in the propagators. 
The so-called fixed-width scheme~\cite{ddrw,gauge} is 
adopted in {\tt WRAP}. Actually, as shown in ref.~\cite{ddrw}, 
the fixed-width scheme, even if it violates $SU(2)$ gauge invariance,
is a reliable $U(1)$ gauge-restoring method and 
is able to guarantee predictions for 
$e^+ e^- \to 4f + \gamma$ processes 
in good numerical agreement with a scheme preserving all the 
relevant Ward identities, such as the complex-mass scheme~\cite{ddrw}.

\subsection{Tree-level matrix element}

As already mentioned, the present work is based on the calculation 
of the {\it fully massive} Born matrix element of $e^+ e^- \to 4f + \gamma$ processes.
The exact matrix 
elements for CC and NC $e^+ e^- \to 4f + \gamma$ processes are 
available in {\tt WRAP}.
The calculation is performed by using {\tt ALPHA}~\cite{alpha}, an 
iterative algorithm for the automatic evaluation of tree-level scattering
amplitudes without using Feynman graphs (see ref.~\cite{mp} for a 
review of the method and of recent phenomenological 
applications). For the processes under consideration, 
a completely numerical approach turns out to
be particularly convenient not only for the very large amount of 
contributing Feynman diagrams, but also because the calculation can be performed 
in the presence of fermion masses without any additional complication.
This is of special importance for $4f + \gamma$ final states 
involving muons, where the separation angle between muon and photon can be 
realistically set to zero, and a calculation taking into account of the finite 
muon mass is mandatory, to avoid collinear singularities. 

\subsection{Anomalous gauge couplings}

Information about the structure of TGC and QGC 
can be obtained by the analysis of $4f + \gamma$ production processes. 
In particular, CC radiative $4f$ processes, although 
characterized by a lower statistics,  
are potentially a complementary channel to the $4f$ final 
states in order to test the effect of TGC, because of the
larger amount of diagrams involving trilinear gauge interactions. 
More importantly, $4f + \gamma$  processes are interesting in 
order to put bounds on deviations from standard quartic gauge couplings.
In the following, the theoretical details of the parameterization adopted 
in order to keep under under control 
this important phenomenological issue are described.

\subsubsection{Trilinear anomalous gauge couplings}     

It is possible to take into account the effect of 
charged TGC (anomalous and not) 
by means of the following lagrangian~\cite{agc,agc2}: 
\begin{equation}
\begin{array}{ll}
i{L}_{TGC}\hspace{1pt} & = \hspace{1pt}g_{WWV}
\left[ \right.
g_1^VV^{\mu}\left(\hspace{2pt}W^-_{\mu\nu}W^{+\nu}
-\hspace{2pt}W^+_{\mu\nu}W^{-\nu}\hspace{2pt}\right)\\
&+\kappa_V W^+_{\mu}W^-_{\nu}V^{\mu\nu}+
\frac{\lambda_V}{m^2_W}V^{\mu\nu}W_{\mu}^{+\rho}W_{\rho\nu}^
-\hspace{2pt}\\
&+\hspace{2pt}ig_5^V\varepsilon_{\mu\nu\rho\sigma}
\left(\left(\partial^{\rho}W^{-\mu}\right)W^{+\nu}
-W^{-\mu}\left(\partial^{\rho}W^{+\nu}\right)\right)V^{\sigma} \left. \right]\\
& V = \gamma, Z, 
\end{array}
\label{tri}
\end{equation}   
which represents the most general lagrangian describing trilinear
$WWV$ gauge interactions, with the exception of the operators violating 
$C$, $P$ and $CP$ simmetries. ${L}_{TGC}$ has been 
implemented in {\tt ALPHA}, and the presence
of anomalous couplings can be studied, as 
done at LEP~\cite{sj}, by using the relations~\cite{agc2,yel} 
\begin{equation}
\Delta\kappa_{\gamma} = -\frac{c^2_W}{s^2_W}(\Delta\kappa_Z-\Delta
g_1^Z)\hspace{1cm}\lambda_Z = \lambda_{\gamma}\equiv \lambda,
\end{equation}
where $\Delta\kappa_V = \kappa_V-1$ and $\Delta g_1^Z = g_1^Z-1$. 
The Standard Model (SM) Lagrangian is recovered for $g_1^V = k_V = 1$, $\lambda = 0$, 
$g_5^V = 0$. Triple anomalous neutral gauge couplings, considered 
in ref.~\cite{r} and looked for at LEP in $e^+ e^- \to Z\gamma, ZZ$ 
processes~\cite{cmo}, are not presently taken into account.

\subsubsection{Quartic anomalous gauge couplings}     

Quartic gauge couplings involving at least one photon 
are analyzed at LEP~\cite{ss}. In particular, $W^+W^-\gamma\gamma$ and 
$W^+W^- Z\gamma$ vertices are probed in 
$WW\gamma \to 4f +\gamma$ and $\nu\bar\nu\gamma\gamma$ 
final states~\cite{l3quad,opalqagc}, while 
$e^+ e^- \to Z\gamma\gamma$ processes~\cite{l3q} 
are investigated to put bounds on the $ZZ\gamma\gamma$ vertex, which is
a gauge interaction not predicted by SM at tree level. 
 In the present work the operators 
considered in ref.~\cite{bela} for genuine anomalous
quartic couplings containing at least one photon, namely 
$W^+W^- Z\gamma$, $W^+W^- \gamma\gamma$ and $ZZ\gamma\gamma$ 
vertices, have been implemented in {\tt ALPHA}, upgrading the 
version used in ref.~\cite{gange} for the analysis of QAGC 
in six-fermion final states 
at the energies of future linear colliders. 
The implemented lagrangian include all the 
relevant six-dimensional operators and reads as follows 
\begin{eqnarray}
L_{QGC} =&& W_1 + W_2 + Z_1 +  Z_2 \nonumber\\
   && W_0^Z +  W_c^Z + W_1^Z + 
    W_2^Z +  W_3^Z .  
\label{wwzgan1}
\end{eqnarray}
In the above equation the Lorentz structure of the operators is given by:
\begin{equation}
\begin{array}{llllll}
W_1  = & a_{w1} F_{\mu\nu}F^{\mu\nu}W^+_{\rho}W^{-\rho} \\
W_2 = & a_{w2} F_{\mu\nu}F^{\mu\rho}W^{+\nu}W^-_{\rho}+ {\rm h.c.}\\   
Z_1  = & a_{z1} F_{\mu\nu}F^{\mu\nu}Z_{\rho}Z^{\rho} \\
Z_2 = & a_{z2} F_{\mu\nu}F^{\mu\rho}Z^{\nu}Z_{\rho}\\   
W_0^{Z} = & a_{wz0} F_{\mu\nu}Z^{\mu\nu}W^+_{\rho}W^{-\rho} \\ 
W_c^{Z} = & a_{wzc} F_{\mu\nu}Z^{\mu\rho}W^{+\nu}W^-_{\rho}+ {\rm h.c.}\\
W_1^{Z} = & a_{wz1} F_{\mu\nu}W^{+\mu\nu}Z^{\rho}W^-_{\rho}+ {\rm h.c.}\\
W_2^{Z} = & a_{wz2} F_{\mu\nu}W^{+\mu\rho}Z^{\nu}W^-_{\rho}+ 
{\rm h.c.}\\ 
W_3^{Z} = & a_{wz3} F_{\mu\nu}W^{+\mu\rho}Z_{\rho}W^{-\nu}+ {\rm h.c. } \, ,
                              &&
\end{array}
\label{wwzgan2}
\end{equation}
where the $a_i$ are coefficients of dimension $M^{-2}$. It is worth 
noticing that, by imposing appropriate relations between the 
$a_i$'s, symmetry properties, such as for instance $SU(2)_c$ custodial 
symmetry or $SU(2) \times U(1)$ gauge invariance, can be guaranteed, 
as shown in ref.~\cite{bela}. In the parameterization adopted in ref.~\cite{bela} the 
$a_i$ are real coefficients whose explicit expression can be 
directly read off from the corresponding operator structure 
of ref.~\cite{bela} itself. In particular, the coefficients $a_0$ 
and $a_c$, originally introduced in ref.~\cite{bb} and related to the 
$WW\gamma\gamma$ and $ZZ\gamma\gamma$ structure, can be obtained from 
the above $a_i$
coefficients by means of the following relations
\begin{equation}
\begin{array}{lcl}
a_{w1} = &-\frac{e^2}{8 \Lambda^2}&a_0\\
a_{z1} = &-\frac{e^2}{16 \cos^2\theta_w \Lambda^2}&a_0\vspace{7pt}\\
a_{w2} = &-\frac{e^2}{16 \Lambda^2}&a_c\\
a_{z2} = &-\frac{e^2}{16 \cos^2\theta_w \Lambda^2}&a_c\ ,
\end{array}
\end{equation}
where $\Lambda$ represents a scale of new physics.
As far as $WWZ\gamma$ vertex is concerned, an additional structure has been 
proposed in the literature~\cite{stirlingwer,an}, 
whose expression can be derived 
from the above $a_i$ coefficients by means of the following relations 
\begin{equation}
\begin{array}{l}
a_{wzc} = i \frac{e^2}{16 \cos\theta_w  \Lambda^2}a_n \\
a_{wz2} = i \frac{e^2}{16 \cos\theta_w  \Lambda^2}a_n  \\
a_{wz3} = -i \frac{e^2}{16 \cos\theta_w  \Lambda^2}a_n \, .
\end{array}
\end{equation}
On the experimental side, bounds on $a_0$, $a_c$ and $a_n$  
couplings are quoted by LEP collaborations via the 
analysis of $WW\gamma$ and $\nu\bar\nu\gamma\gamma$ 
final states~\cite{l3quad,opalqagc,ss}. 
It is worth noticing, in passing, that, 
thanks to the implementation of the lagrangian of eq.~(\ref{wwzgan1}) 
in the {\tt ALPHA} code, an improved 
version of the Monte Carlo generator {\tt NUNUGPV}~\cite{sfpt,nunu} is also 
available for the study of QAGC in $\nu\bar\nu\gamma\gamma$ 
events. 

The result of the implementation in {\tt ALPHA} has been 
carefully cross-checked by an independent analytical 
calculation of all $V_1 V_2 \to V_3 V_4$ amplitudes, 
with $V_i = \gamma, W, Z$. The check has been performed for all the 
processes obtained from $W^+W^-\to Z\gamma$ scattering, 
by permutating particles.

\subsection{Initial-state radiation}

In order to match the precision of LEP measurements, 
the most important radiative corrections have to 
be considered. Among them, it is well known that 
undetected initial-state radiation (ISR) plays a major r$\hat{{\rm o}}$le.
It can be taken into account in the 
leading log approximation  by using the QED
Structure Function (SF) approach, in terms of
collinear~\cite{ore} or
$p_{\perp}$-dependent SF~\cite{sfpt,nunu}. Following 
recent work done in ref.~\cite{nunu}, both 
 prescriptions are available in {\tt WRAP}, for the reason explained below. 
When ISR is included via collinear SF, the QED corrected 
cross section can be 
written as
\begin{equation}
\sigma^{4f + 1\gamma}_{QED} (s) =
\int\, dx_1dx_2\hspace{2pt}D(x_1,s)D(x_2,s)
\, d \sigma_0^{4f + 1\gamma}(x_1x_2s) ,
\label{sf}
\end{equation}
by convoluting the tree-level cross section with 
electron SF. However, due to the presence of an observed photon in the 
hard-scattering  matrix element, the
inclusion of ISR needs some care. Actually, since 
the collinear SF can be viewed as the result of an integration over
the angular variables of the photon radiation, an overlapping  between
the detected kernel photon and pre-emission photons at
large angle may occur\footnote{The same problem 
is discussed in detail in ref.~\cite{nunu} for the process 
$e^+ e^- \to \nu\bar\nu + n \gamma$. We refer the reader to
ref.~\cite{nunu} for more details 
on the strategy here adopted.}. 
The consequence is that a double
counting takes place if higher-order QED corrections are naively 
included by using collinear SF~\cite{nunu}.

On the other hand, it is expected that the bulk of the correction is well 
estimated by collinear SF, since  the emission of a photon from an
on-shell initial state fermion is almost collinear. However,  
in order to provide a more appropriate 
treatment of photon corrections and give an estimate of the double-counting effect, 
the SF method can be improved by means of 
the use of $p_{\perp}$-dependent SF, i.e. by 
generating angular variables for the ISR photons according to 
 $1/(p\cdot k)$, which is the leading behaviour for radiation
of momentum $k$ emitted by an on-shell fermion 
of momentum $p$. In such a scheme, 
the QED corrected cross section can be calculated as
\begin{eqnarray}
\sigma^{4f + 1\gamma}_{QED} =&& \int d x_1 d x_2 \int_{\Omega_c} d
c_{\gamma}^{(1)} d c_{\gamma}^{(2)} \nonumber \\ \,
&& \tilde{D} (x_1, c_{\gamma}^{(1)}; s) \tilde{D} (x_2, c_{\gamma}^{(2)}; s)
d\sigma^{4f + 1\gamma} ,
\label{pt}
\end{eqnarray}
where $D(x, c_{\gamma};s)$ is the $p_{\perp}$-dependent SF~\cite{sfpt}.
According to eq.~(\ref{pt}), an ``equivalent '' photon 
is generated and accepted as an ISR contribution only if it satisfies a 
rejection algorithm based on the following requirements:
\begin{itemize}
\item the energy of the ``equivalent'' photon is below the energy
threshold for the observed photon, for arbitrary angles; or
\item the ``equivalent'' photon is collinear to a charged particle
(i.e. under the minimum
separation angle required in order to be detected), for arbitrary energies.
\end{itemize}

Within the angular acceptance of the observed photon, the cross section 
is computed by means of the exact matrix element for the 
considered $4f+\gamma$ final state. Therefore, eq.~(\ref{pt}) applies 
to the signature of four fermions plus an isolated hard photon, corrected
by the effect of undetected soft and/or collinear radiation. 
Aiming to obtain a correct evaluation of the size of the double counting effects, 
a limit of the present treatment of undetected radiation 
is that only ISR is actually considered. 
This issue could be addressed in a more complete way by using, for 
example, a QED parton shower approach as developed in 
ref.~\cite{babayaga}, 
in order to describe the radiation 
from all external charged legs, thus including 
the contribution of undetected final-state radiation. 

\section{Numerical results}

The aim of the present section is to give some details on the 
technical precision of {\tt WRAP} and discuss the impact of the effects
due to fermion masses, ISR and AGC on observables of experimental interest.   

\begin{table}
\caption{Comparison between {\tt WRAP} and {\tt RacoonWW} predictions 
for the massless Born cross 
section of NC processes at $\sqrt{s}=190$~GeV. Input parameters and 
cuts as in ref.~\cite{ddrw}.}
\label{ddrwnc}       
\begin{tabular}{llll}
\hline\noalign{\smallskip}
Cross section (fb) &{\tt WRAP} & {\tt RacoonWW}   \\
\noalign{\smallskip}\hline\noalign{\smallskip}
$\mu^+\mu^-\tau^+\tau^-\gamma$ & $6.76 \pm 0.03$ & $6.78 \pm 0.03$ \\
$\mu^+\mu^-\nu_{\tau}\bar{\nu}_{\tau}\gamma$ & $4.248 \pm 0.009$ & $4.259 \pm 0.009$\\
$\mu^+\mu^- u {\bar u}\gamma$ & $12.65 \pm 0.03$ & $12.70 \pm 0.04$ \\
\noalign{\smallskip}\hline
\end{tabular}
\end{table}

In order to test the reliability and the theoretical accuracy 
of the event generators,
a detailed tuned comparison between the predictions of {\tt WRAP} and 
other available programs have been carried out in the 
context of the four-fermion working group of the LEP2 Monte Carlo
workshop at CERN~\cite{lep2mc}. The comparisons, referred to 
integrated cross sections and differential 
distributions of several CC processes, showed perfect 
technical agreement. The comparison is here extended to NC processes, 
as shown in Tab.~\ref{ddrwnc}, between the predictions of {\tt WRAP} 
and {\tt RacoonWW} with input parameters and cuts as in ref.~\cite{ddrw}. As can be 
seen, also for NC final states the agreement is excellent. A further
 comparison between the predictions of {\tt WRAP} and those 
 of ref.~\cite{kj2} is reported in Tab.~\ref{kj2c}, for several 
 cross sections of CC and NC processes, in the presence of finite
 fermion masses and in terms of the same input parameters and 
 cuts as adopted in ref.~\cite{kj2}. Perfect agreement is registered 
 for all the  considered $4f + \gamma$ final states.

\begin{table}
\caption{Comparison between {\tt WRAP} and the predictions 
of ref.~\cite{kj2} for the massive Born cross 
section of CC and NC processes at $\sqrt{s}=190$~GeV. Input parameters and 
cuts as in ref.~\cite{kj2}.}
\label{kj2c}       
\begin{tabular}{llll}
\hline\noalign{\smallskip}
Cross section (fb) &{\tt WRAP} & ref.~\cite{kj2}   \\
\noalign{\smallskip}\hline\noalign{\smallskip}
$u\bar{d}e^-\bar\nu_e\gamma$ & $220.1 \pm 0.5$ & $220.3 \pm 0.7$ \\
$c\bar{s}e^-\bar\nu_e\gamma$ & $217.5 \pm 0.4$ & $218.2 \pm 0.7$ \\
$\mu^+\bar{\nu}_{\mu}e^-\bar\nu_e\gamma$ & $78.6 \pm 0.1$ & $79.0 \pm 0.3$ \\
$\tau^+\bar{\nu}_{\tau}e^-\bar\nu_e\gamma$ & $77.6 \pm 0.2$ & $77.5 \pm 0.2$ \\
$u\bar{d}\mu^-\bar\nu_{\mu}\gamma$ & $213.0 \pm 0.1$ & $213.8 \pm 0.3$ \\
$u\bar{d}\tau^-\bar\nu_{\tau}\gamma$ & $208.7 \pm 0.4$ & $209.3 \pm 0.5$ \\
$\tau^+\bar{\nu}_{\tau}\mu^-\bar\nu_{\mu}\gamma$ & $75.2 \pm 0.1$ & $75.1 \pm 0.2$ \\
$u\bar{d}s\bar{c}\gamma$ & $590.0 \pm 0.6$ & $593 \pm 2$ \\
$\mu^+\mu^-\nu_{\tau}\bar{\nu}_{\tau}\gamma$ & $5.32 \pm 0.02$ & $5.32 \pm 0.03$ \\
$\tau^+\tau^-\mu^+\mu^-\gamma$ & $4.15 \pm 0.02$ & $4.18 \pm 0.02$ \\
$\tau^+\tau^-\nu_{\mu}\bar{\nu}_{\mu}\gamma$ & $3.175 \pm 0.006$ & $3.167 \pm 0.007$ \\
\noalign{\smallskip}\hline
\end{tabular}
\end{table}

The phenomelogical analysis makes use of the following input parameters:

\begin{equation}
\begin{array}{llll}
G_F=1.16637\cdot10^{-5}\hspace{2pt}{\rm GeV}^{-2} \hspace{.5cm} & 
M_Z=91.1867\hspace{2pt} {\rm GeV} \\
M_W=80.35 \hspace{2pt}{\rm GeV}
& \sin^2\theta_w = 1 - M_W^2/M_Z^2\\
\Gamma_Z=2.49471 \hspace{2pt} {\rm GeV} & \Gamma_W=2.04277 \hspace{2pt} 
{\rm GeV} \\
m_{\mu} = 0.10565839 \hspace{2pt} {\rm GeV} & m_s = 0.15 \hspace{2pt} 
{\rm GeV}\\
m_{c} = 1.55 \hspace{2pt} {\rm GeV}
\end{array} 
\label{param}
\end{equation}

The form used for the propagator of
the massive gauge bosons is, according to the fixed-width scheme, 
$\sim 1/(p^2 - M^2 + i\Gamma M)$. 
The processes considered are the 
radiative semi-leptonic final states of the kind 
$e^+e^-\to l^+ {\nu}_l q {\bar q^{'}} \gamma$. 
The cuts adopted are:
\begin{equation}
\begin{array}{l}
\vert \cos\theta_{\gamma}\vert \le 0.985; E_{\gamma}
\ge
1\hspace{2pt} {\rm GeV};\\
\vert \cos\theta_{l}\vert \le 0.985; E_{l}\ge
5\hspace{2pt}{\rm GeV};\\
\theta_{\gamma-f} \ge 5^{\circ};\\
M_{q{\bar q}^{'}}\ge 10\hspace{2pt}{\rm GeV} ,
\end{array}
\label{cuts}
\end{equation}
where ${\theta_{\gamma}}({_l})$ is the photon(lepton) scattering angle, 
${E_{\gamma}}({_l})$ is the photon(lepton) energy, 
$\theta_{\gamma-f}$ is the angular separation between photon and 
final charged fermions, and $M_{q{\bar q}^{'}}$ is the
$q{\bar q}^{'}$ invariant mass.

\begin{table}
\caption{Comparison between massive and massless Born cross section for the 
final state $\mu^+ {\nu}_{\mu} \bar{c} s + \gamma$ at $\sqrt{s}=200$~GeV. 
$\theta_{\gamma -f}$ is the minimum separation angle between the photon and 
final state charged fermions; other cuts as 
in eq.~(\ref{cuts}). The first line refers to the
massive case, the second one to the massless approximation.}
\label{masstable}       
\begin{tabular}{ll}
\hline\noalign{\smallskip}
$\theta_{\gamma -f}$ (deg) & Cross Section (fb)  \\
\noalign{\smallskip}\hline\noalign{\smallskip}
$5^{\circ}$              & $74.294 \pm 0.029$\\
                         & $75.732 \pm 0.022$ \\
\noalign{\smallskip}\hline
$1^{\circ}$              & $~\,93.764 \pm 0.037$ \\           
                         & $100.446 \pm 0.037$ \\ 
\noalign{\smallskip}\hline
\end{tabular}
\end{table}

\begin{table}
\caption{Comparison between massive and massless Born cross sections for the 
final state $\mu^+ {\nu}_{\mu} \bar{c} s + \gamma$ at $\sqrt{s}=200$~GeV.
$\theta_{\gamma -f}$, with $f = q, \mu$ is the minimum separation angle 
between the photon and 
final state charged fermions; other cuts as 
in eq.~(\ref{cuts}). In the third column, the first 
result refers to the
massive case, and the second one to the massless case. The relative difference 
is shown in the last column. See also ref.~\cite{lep2mc}.}
\label{zerodeg}       
\begin{tabular}{llll}
\hline\noalign{\smallskip}
$\vartheta_{\gamma -q}$ (deg) & $\vartheta_{\gamma -\mu}$ (deg)   & 
Cross Section (fb) & \, \,  $\delta$ (\%) \\
\noalign{\smallskip}\hline\noalign{\smallskip}
$5^{\circ}$     & $1.0^{\circ}$      & $90.157 \pm 0.036$  & $1.92 \pm 0.08$ \\
                &                    & $91.903 \pm 0.035$  & \\
\noalign{\smallskip}\hline
$5^{\circ}$     & $0.1^{\circ}$      & $104.777 \pm 0.046$ & $9.31 \pm 0.09$ \\
                &                    & $115.004 \pm 0.044$ &   \\  
\noalign{\smallskip}\hline
 $5^{\circ}$    & $0.0^{\circ}$      & $105.438 \pm 0.045$ & \\ 
\noalign{\smallskip}\hline
\end{tabular}
\end{table}

\begin{figure*}
\begin{center}
\includegraphics[bb= 25 230 525 585,scale=.7]{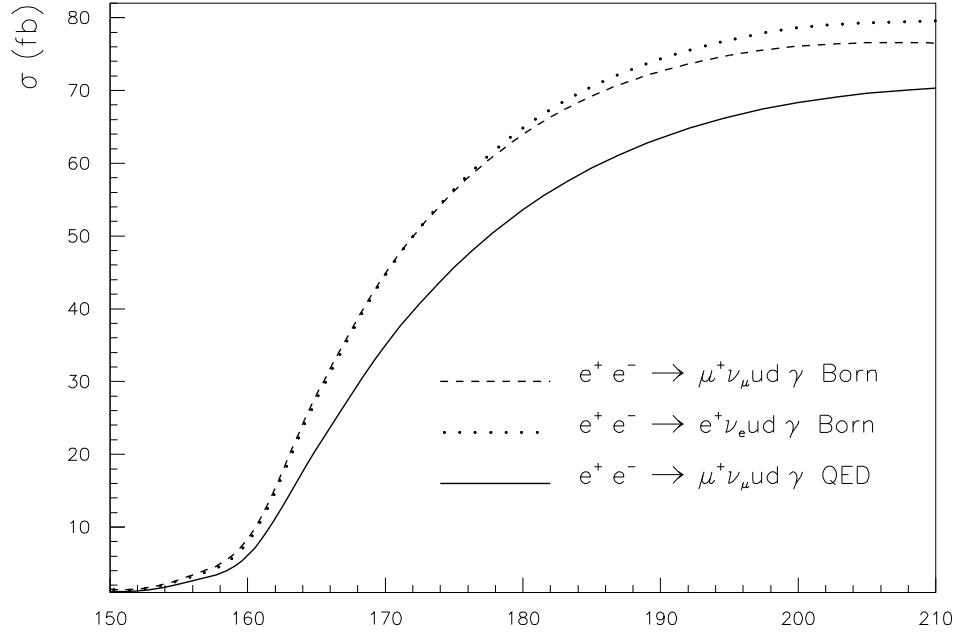}
\caption{Cross section for the semi-leptonic processes $e^+e^-\to
l^+ {\nu}_{l}\bar{u}d\gamma$ with $l^+=\mu^+$ (dashed line) and 
$l^+=e^+$ (dotted line). The solid line shows 
the QED corrected cross section via collinear SF for the  
$\mu^+{\nu}_{\mu}\bar{u}d\gamma$ final state. 
See also ref.~\cite{lep2mc}.}
\label{fig4_epj}
\end{center}
\end{figure*}
\noindent

\begin{figure*}
\begin{center}
\includegraphics[bb= 25 230 525 585,scale=.7]{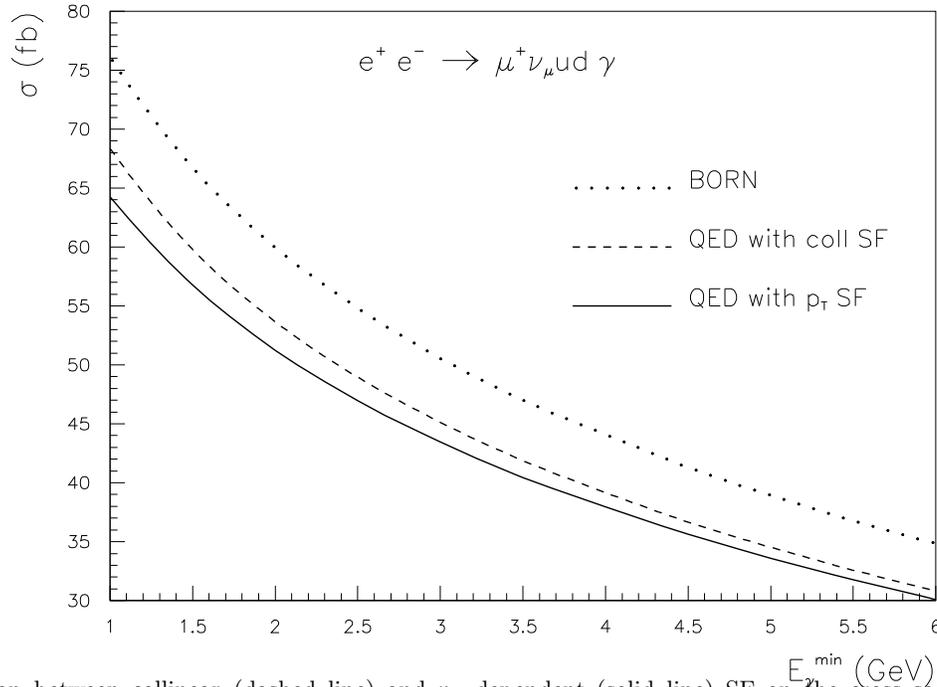}
\caption{Comparison between collinear (dashed line) and $p_{\perp}$-dependent
(solid line) SF 
 on the cross section of the process $e^+ e^- 
 \to \mu^+{\nu}_{\mu}\bar{u}d\gamma$, 
as a function of the energy threshold of the visible photon
$E_{\gamma}^{min}$. The dotted line is the Born prediction. 
See also ref.~\cite{lep2mc}.}
\label{fig5_epj}
\end{center}
\end{figure*}

Table~\ref{masstable} shows the effect of fermion masses on integrated 
cross sections at $\sqrt{s}=200$~GeV 
for two different photon-fermion separation angles. In the first row, 
the angular separation 
 $\theta_{\gamma -f}$ between photon and all charged final state fermions 
is fixed at $5^{\circ}$, while in the second row  
$\theta_{\gamma -f}= 1^{\circ}$. As expected, 
the relative difference between the massless and massive cross section increases,
going from $2\%$ of the first row to the $7\%$ of the second row, because
of the importance of fermion mass contributions when the photon approaches 
the collinear region around an on--shell charged particle. 
In the case of a final state containing a muon, the separation angle 
$\vartheta_{\gamma -f}$ can be realistically  set to zero, because of 
different behaviour of photons and muons in the experimental
apparatus. Table~\ref{zerodeg} shows the difference between massive and 
massless cross section, with 
the minimal separation between 
quarks and photon fixed at $5^{\circ}$ and progressively relaxing the separation cut 
between muon and photon. It can be 
seen that the massless calculation is still reliable for $1^{\circ}$ of minimum
separation, being the relative difference around $2\%$, but it becomes 
inadequate when the separation falls at some fraction
of degree, the relative difference being of the order 
of $10\%$. Therefore, in particularly stringent experimental conditions,
only a massive $4f+\gamma$ calculation can provide 
reliable predictions in the presence of muons in the final state.

Figure~\ref{fig4_epj} shows the line-shape of the cross sections 
of the radiative semi-leptonic processes 
$\mu^+ {\nu}_{\mu} \bar{u} d + \gamma$ and 
$e^+ {\nu}_{e} \bar{u} d + \gamma$, as a function of the c.m. energy 
in  the LEP2 range. The QED corrected cross section via collinear SF 
for the $\mu^+ {\nu}_{\mu} \bar{u} d + \gamma$  
final state is also plotted. The comparison shows that 
the contribution due to the additional $t$-channel
diagrams present in the $e^+ {\nu}_{e} \bar{u} d + \gamma$ 
final state is not particularly relevant for the adopted 
selection criteria, small being the differences between the cross 
sections of the two processes. Concerning ISR in the 
strictly collinear approximation, its 
impact on the cross section is at the level of $10-15\%$, which
is a phenomenologically relevant effect in the light of  
the LEP experimental accuracy. It is worth noticing that this 
result, obtained by means of a standard treatment of ISR 
as tipically adopted in the experimental analysis of QAGC 
in radiative events at LEP~\cite{l3quad,opalqagc}, just provides 
the bulk of the effect due to ISR but it is affected, 
as previously discussed and quantified below, by a 
double counting because of the presence of a radiative 
process as hard-scattering reaction.
 
The contribution of initial-state photon radiation 
is also shown in Figs.~\ref{fig5_epj}, 
 as a function of the threshold energy
$E^{min}_{\gamma}$ of the observed photon. It can be 
noticed that the reduction factor due to collinear ISR
is around 12-13\%, almost independent of the photon 
detection threshold. However, as previously discussed,  
ISR in the collinear SF scheme introduces a double
counting effect when the pre-emission ``equivalent'' photon enters the
phase-space region of the kernel photon. In order to get an estimate 
of this overlapping contribution, the 
comparison of the corrections due to the collinear SF and $p_{\perp}$-dependent 
SF is shown. Is is observed that the two prescriptions can 
differ at 5\% level for $E^{min}_{\gamma}$ close to 1-2 GeV, while 
the becomes smaller and smaller as $E^{min}_{\gamma}$ increases. Numerical
investigation points out that, as expected, the discrepancy between 
collinear and $p_{\perp}$-dependent SF is larger near the soft and collinear 
region and at the level of some per cent, thus yielding an 
estimate of the size of the double-counting effect at the 
level of ISR. Therefore, in the presence of particularly
stringent experimental constraints sensitive to the soft and
collinear emission, precise predictions demand a 
treatment of ISR able to keep under control the 
transverse degrees of freedom of photon radiation. 

\begin{figure*}
\begin{center}
\includegraphics[bb= 15 215 540 590,scale=.7]{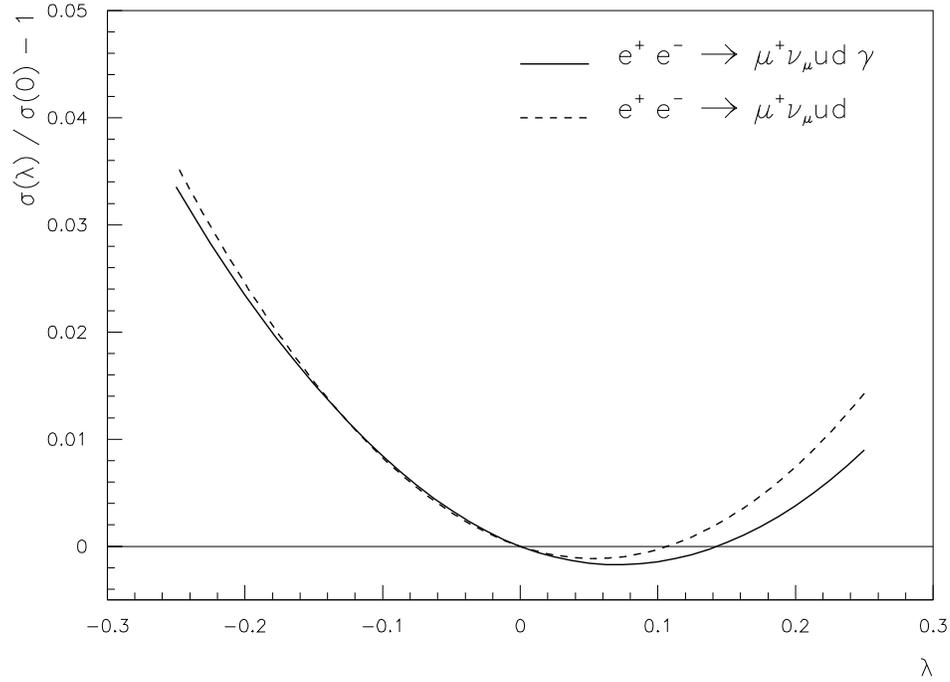}
\caption{The relative effect of TAGC $\lambda$ on the cross section
of the radiative process $e^+e^-\to \mu^+\nu_{\mu}\bar{u}d \gamma$ (solid line) and the
corresponding $4f$ final state (dashed line). }
\label{fig6_epj}
\end{center}
\end{figure*}

\begin{figure*}
\begin{center}
\includegraphics[bb= 15 215 540 590,scale=.7]{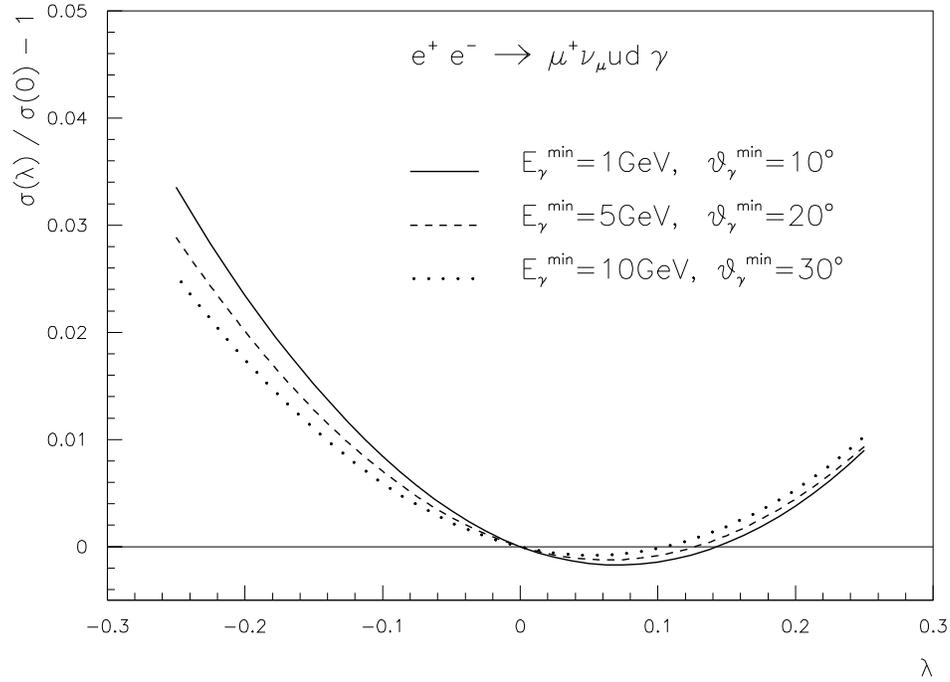}
\caption{The relative effect of TAGC $\lambda$ on the cross section
of the radiative process $e^+e^-\to \mu^+\nu_{\mu}\bar{u}d \gamma$ 
for different photon 
selection criteria.}
\label{fig7_epj}
\end{center}
\end{figure*}

\begin{figure*}
\begin{center}
\includegraphics[bb= 15 215 540 590,scale=.6]{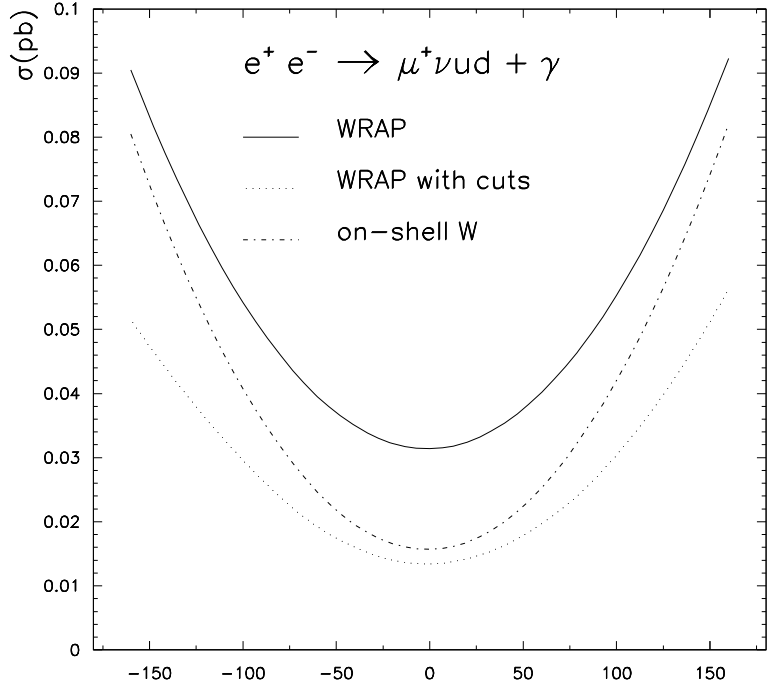}
\vskip 24pt
\caption{The effect of the QAGC $k_0^W$ at $\sqrt{s} = 200$~GeV 
on the absolute cross section
for the process $e^+e^-\to \mu^+\nu_{\mu}\bar{u}d + \gamma$,  
with $\Lambda = M_W$. 
The solid line is obtained by means of the full calculation of {\tt WRAP}, 
the dash-dotted one with the real $W W \gamma$ approximation, and the 
dotted line refers to the calculation of {\tt WRAP} with the additional 
cuts 75~GeV $\leq M(\mu^+ \nu_{\mu}), M({\bar u} d) \leq $ 85 GeV.}
\label{fig8_epj}
\end{center}
\end{figure*}

\begin{figure*}
\begin{center}
\includegraphics[bb= 15 215 540 590,scale=.6]{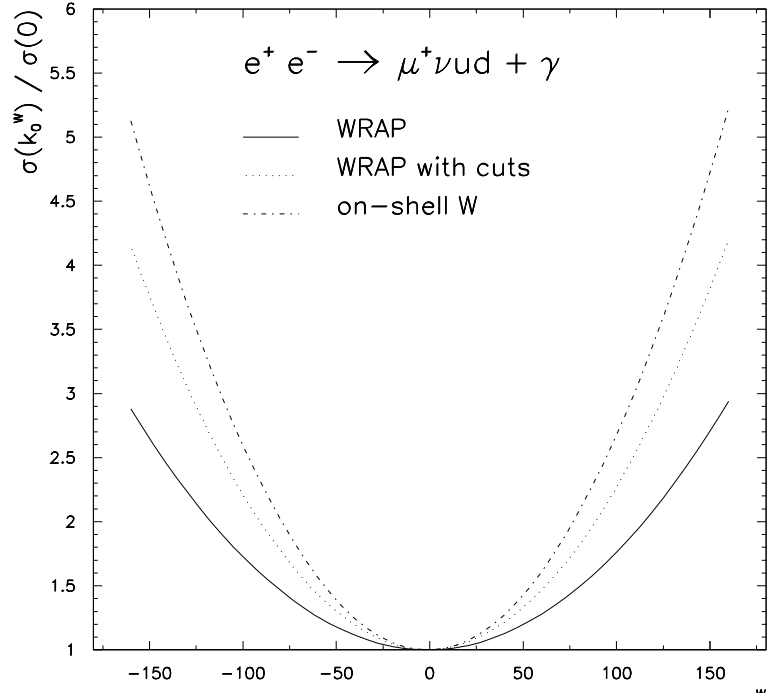}
\vskip 24pt
\caption{The ratio between the cross section in the presence of a QAGC $k_0^W$ 
and the SM cross section, as a function of $k_0^W$, 
for the three different cases as in Fig.~\ref{fig8_epj}.}
\label{fig9_epj}
\end{center}
\end{figure*}

\begin{figure*}
\begin{center}
\includegraphics[bb= 30 196 550 615,scale=.6]{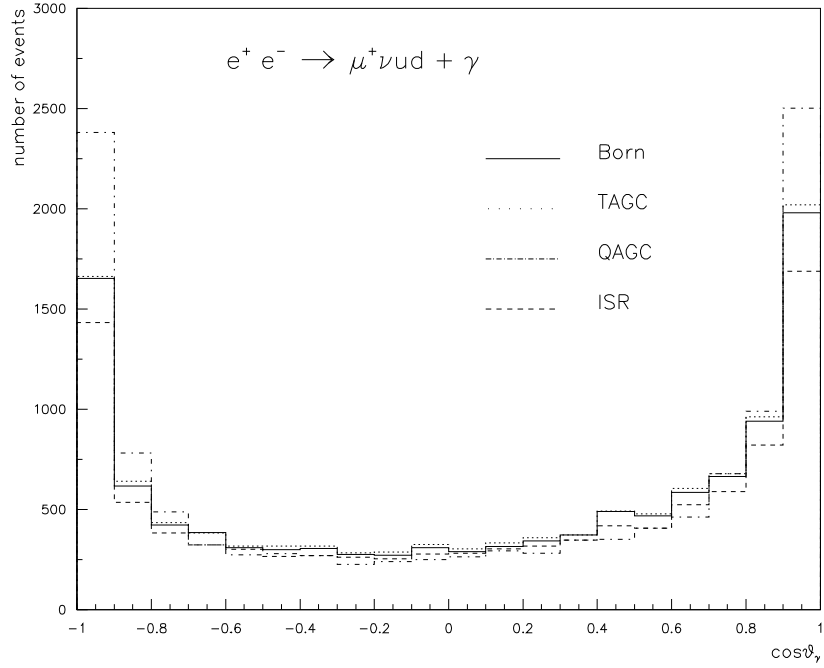}
\caption{The $\cos\theta_{\gamma}$ distribution for the process 
$e^+ e^- \to \mu^+ \nu_{\mu}\bar{u}d\gamma$ at $\sqrt{s} = 192$~GeV. 
The Born approximation (solid line),the QED corrected calculation (dashed 
line), the predictions for $\lambda=-0.25$ (dotted line) 
and the ones for $k_0^W / \Lambda^2 = 0.01$ (dash--dotted line) are shown. 
Cuts as in~(\ref{cuts}).}
\label{fig10_epj}
\end{center}
\end{figure*}

\begin{figure*}
\begin{center}
\includegraphics[bb= 30 196 550 615,scale=.6]{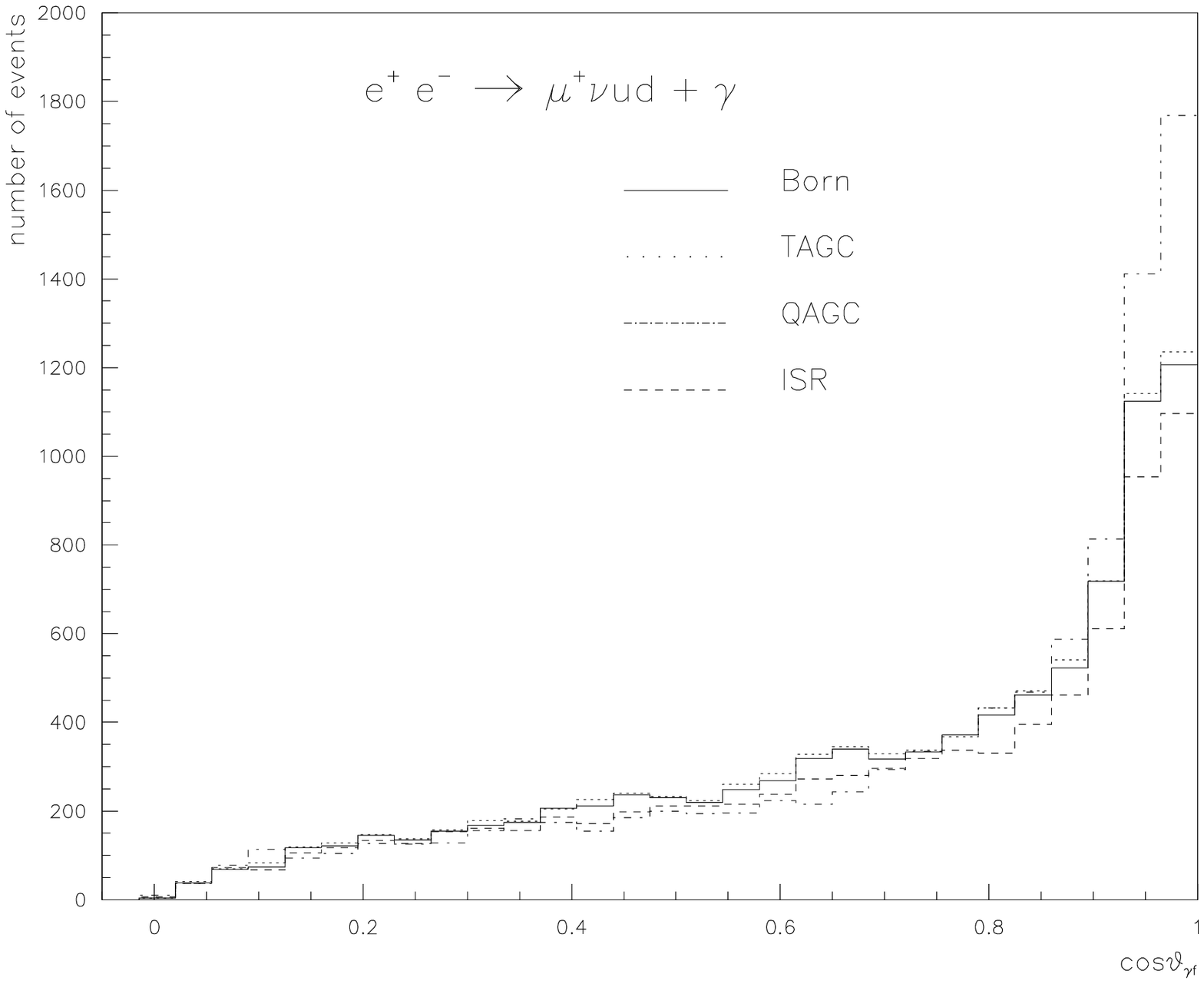}
\caption{The 
$\cos\theta_{\gamma f}$ distribution for the process 
$e^+ e^- \to \mu^+ \nu_{\mu}\bar{u}d\gamma$ at $\sqrt{s} = 192$~GeV, where
$\theta_{\gamma f}$ is the angle between the photon and the nearest
 charged particle. The samples are the same of Fig.~\ref{fig10_epj}.}
\label{fig11_epj}
\end{center}
\end{figure*}

\begin{figure*}
\begin{center}
\includegraphics[bb= 30 196 550 615,scale=.6]{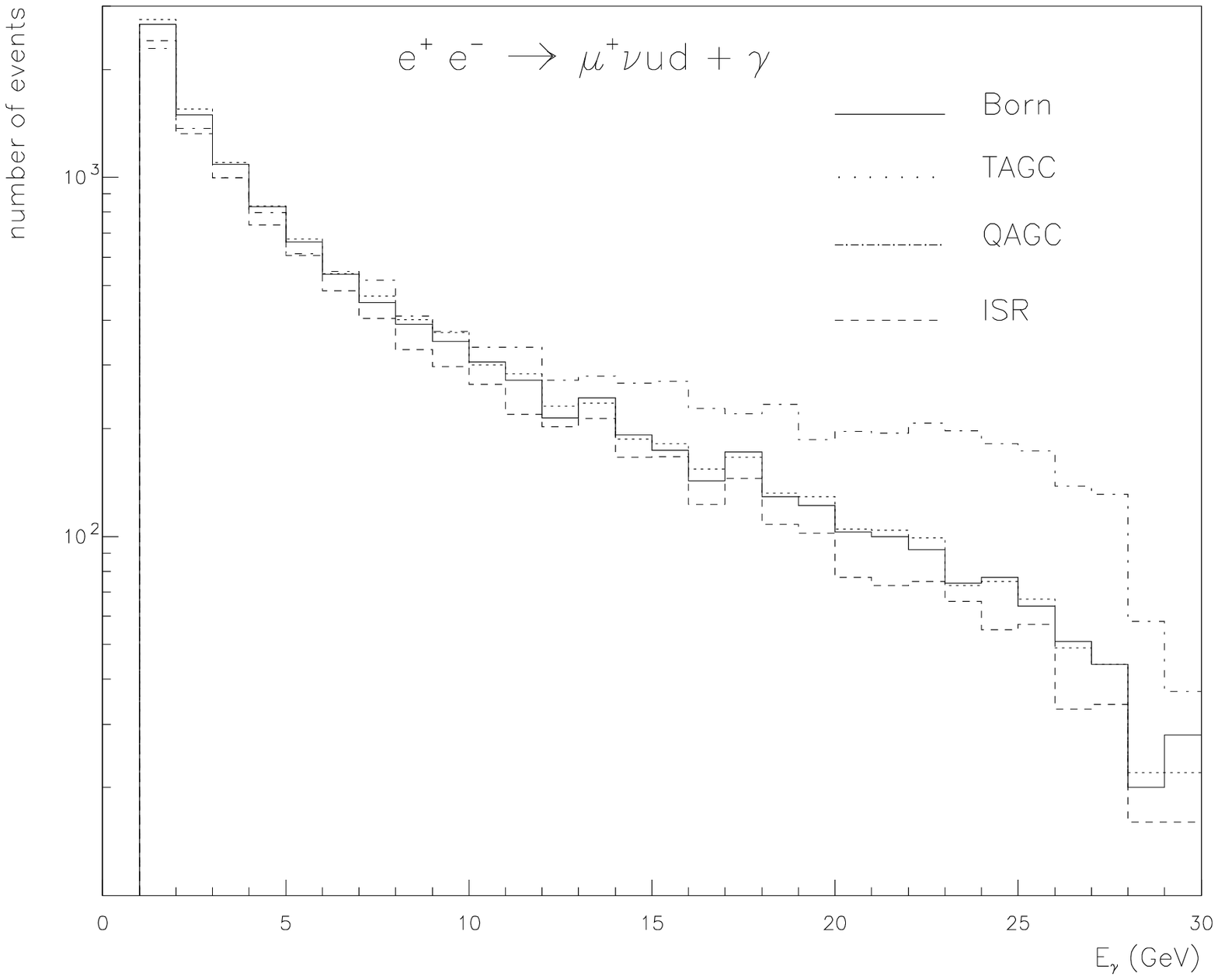}
\caption{The $E_{\gamma}$ distribution for the process $e^+e^-\to
\mu^+\nu_{\mu}\bar{u}d\gamma$ at $\sqrt{s}=192$~GeV for the same sample of
events as in Fig.~\ref{fig10_epj}.}
\label{fig12_epj}
\end{center}
\end{figure*}

\begin{figure*}
\begin{center}
\includegraphics[bb= 30 196 550 615,scale=.6]{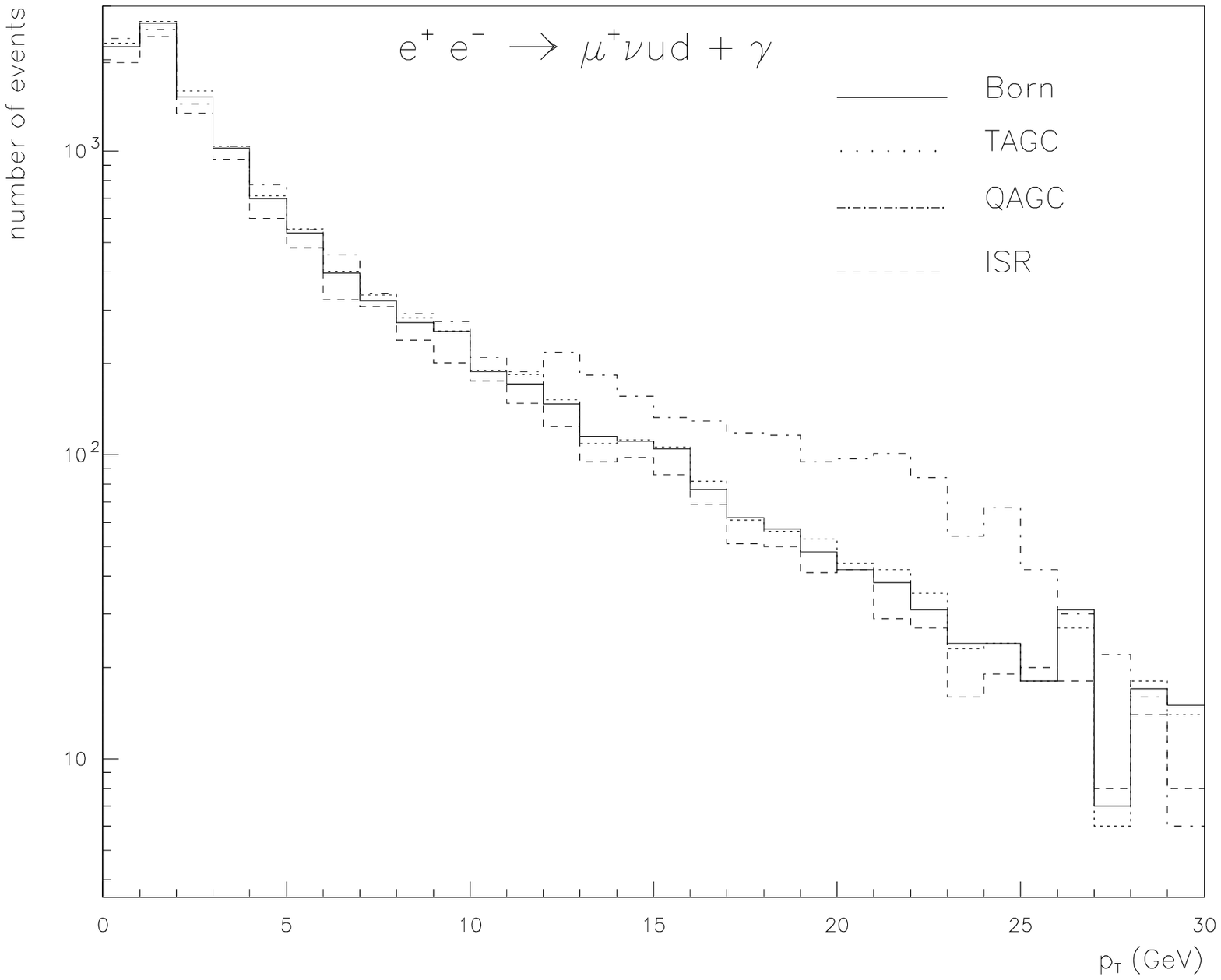}
\caption{The distribution of the transverse momentum of the
visible photon for the process $e^+e^-\to
\mu^+\nu_{\mu}\bar{u}d\gamma$ at $\sqrt{s}=192$~GeV. The events samples are the
same of Fig.~\ref{fig10_epj}.}
\label{fig13_epj}
\end{center}
\end{figure*}

Let us come to the discussion of the effects due to
(a sample of) AGC. Both integrated cross sections 
(Figs.~\ref{fig6_epj}-\ref{fig9_epj}) and differential distributions 
(Figs.~\ref{fig10_epj}-\ref{fig13_epj}) are considered. 
In Figs.~\ref{fig6_epj}-\ref{fig7_epj} the (relative) effect of the TAGC $\lambda$ 
on the $e^+e^-\to \mu^+\nu_{\mu}\bar{u}d\gamma$ cross section is examined, by 
plotting the relative difference
between the cross section in the presence of 
a non-vanishing $\lambda$ coupling and the SM cross section
($\lambda = 0$ ), as a function of the $\lambda$ value at $\sqrt{s} = 192$~GeV.
Fig.~\ref{fig6_epj} shows a comparison of the effect of the 
$\lambda$ coupling on the radiative $\mu^+\nu_{\mu}\bar{u}d + \gamma$ 
process and the corresponding $4f$ final state, which, as already remarked, differ
for their content of trilinear gauge interactions. The numerical 
results for the $4f$ process have been obtained by means 
of the program {\tt WWGENPV}~\cite{wwgenpv}. 
For the considered $\lambda$ values,
the relative contribution is almost the same on the two 
processes (obviously the cross sections are
quite different), giving a difference at $2-3\%$ level only for 
extreme $\lambda$ values. Therefore, in the presence of standard 
cuts on the observed photon, trilinear gauge interactions 
due to $W$ radiation in radiative $4f$ processes doesn't 
enhance the sensitivity to TAGC with respect to a pure $4f$ 
final state. This conclusion is further corroborated by the 
results shown in Fig.~\ref{fig7_epj}, where the effect of the
$\lambda$ coupling is studied for different photon cuts,  
with the aim of suppressing the mostly collinear fermion radiation by imposing
more and more severe cuts on the detected photon. By comparing 
the relative deviations shown in Fig.~\ref{fig6_epj} and Fig.~\ref{fig7_epj}, 
one can conclude that in radiative $4f$ processes $W$
radiation can be hardly disentangled from the radiation off fermions, 
being the observed deviations almost at the same level for all the set of cuts 
considered. 

The effect of the QAGC $k_0^W$, as defined in ref.~\cite{bela}, 
is shown in Figs.~\ref{fig8_epj}-\ref{fig9_epj} for the process 
$\mu^+ \nu_{\mu} {\bar u} d \gamma$, as a 
function of the parameter $k_0^W$ at $\sqrt{s} = 200$~GeV. 
For the scale of new physics $\Lambda$, the value 
$\Lambda = M_W$ is used, as conventionally 
done in the literature. Absolute and relative 
effects are shown in Fig.~\ref{fig8_epj} and Fig.~\ref{fig9_epj}, 
respectively. In terms of the coefficients $a_i$ of 
eq.~(\ref{wwzgan2}) the $k_0^W$ coupling can be expressed as: 
\begin{equation}
\begin{array}{ll}
a_{w1} &= \frac{-e^2g^2}{2\Lambda^2} k_0^W  \\
a_{wz0} &= \frac{-e^2g^2}{\Lambda^2} \frac{\cos\theta_w}{\sin\theta_w} k_0^W.
\end{array}
\nonumber
\end{equation}
The solid line refers to the 
complete $4f + \gamma$ calculation of {\tt WRAP} with input parameters and 
photon cuts as used in ref.~\cite{bela} and the additional cuts on 
fermions as given by eq.~(\ref{cuts}). In order to compare with 
the results of ref.~\cite{bela}, the dash-dotted line has been obtained by means of 
a calculation of the process $e^+ e^- \to W^+ W^- \gamma$ 
(performed independently and in 
agreement with the results of ref.~\cite{bela}), by taking into account 
the suitable branching ratios of the $W$ bosons decaying into $\mu^+ \nu_{\mu}$ and 
${\bar u} d$ pairs. The dotted line is the prediction 
as obtained by {\tt WRAP}, 
with additional cuts on the invariant masses of the two fermionic pairs 
constrained within 75~GeV and 85~GeV, in order to enhance, as 
much as possible, the contribution of diagrams with two resonant $W$ bosons. 
It can be clearly noticed that, even in the 
presence of cuts on the invariant masses of the decay products, 
the complete $4f + \gamma$ calculation differ from the prediction of 
the $WW\gamma$ approximation, thus proving the importance 
of a full calculation for the extraction of meaningful
limits on QAGC.

In Figs.~\ref{fig10_epj}-\ref{fig13_epj} the most important photonic distributions are 
displayed using the code as event generator with the cuts 
of eq.~(\ref{cuts}) at a typical LEP2 energy
 $\sqrt{s} = 192$~GeV. In each plot, the SM Born 
and the QED corrected predictions are compared with those obtained 
in the presence of AGC. The values used for the anomalous couplings are:
$\lambda = -0.25$ and $k_0^W / \Lambda^2 = 0.01$. 
For the sake of comparison, all the data sample 
are normalized to the same luminosity. The $\cos\theta_{\gamma}$ distribution and 
the distribution of the cosine of the angle between the 
photon and the nearest charged particle are shown in 
Fig.~\ref{fig10_epj} and Fig.~\ref{fig11_epj}, respectively. Typical 
peaking behaviour in the close-to-collinear regions is clearly
registered. In such regions, a particularly significative 
impact of the QACG $k_0^W$ is also observed.  
Fig.~\ref{fig12_epj} and Fig.~\ref{fig13_epj} refer to the $E_{\gamma}$ and 
transverse photon momentum $p_\perp$ distribution, respectively, showing 
the characteristic infrared peak. As already noticed in ref.~\cite{bela}, 
these observables turn out to be particularly sensitive 
to the presence of a QAGC 
in the region of high energy and $p_\perp$, being the operator involved 
of derivative type with respect to the photon field. 
In all the considered distributions, ISR introduce sizeable effects 
if compared with the deviations due to anomalous couplings.

As far as the parameterization of QAGC in terms of 
$a_0,a_c,a_n$ parameters is concerned, numerical results 
are shown in Tab.~\ref{qagc200} and Tab.~\ref{qagc500}, 
at $\sqrt{s}=200$~GeV and $\sqrt{s}=500$~GeV, respectively. The cross 
sections in the presence of non-vanishing anomalous couplings
are compared with the pure SM predictions. By looking at 
Tab.~\ref{qagc200} and Tab.~\ref{qagc500}, it can be noticed that
the sensitivity of the $4f$+$\gamma$ processes to QAGC is much
higher at the energies of a future LC than at LEP2, as a priori
expected and already noticed in the literature for the 
$WW\gamma$ process~\cite{stirlingwer,bela}.

\begin{table}
\caption{Effect of the QAGC $a_0/\Lambda^2, 
a_c/\Lambda^2, a_n/\Lambda^2$, with $\Lambda = M_W$, on the cross section of 
the process $e^+ e^- \to u \bar{d} \mu^- \bar{\nu}_{\mu} \gamma$ at 
$\sqrt{s} = 200$~GeV. }
\label{qagc200}       
\begin{tabular}{llll}
\hline\noalign{\smallskip}
QAGC &  Cross section (fb) \\
\noalign{\smallskip}\hline\noalign{\smallskip}
Standard Model & $76.0 \pm 0.1$ \\
$a_0/\Lambda^2 = -0.01 $ &   $77.0 \pm 0.1$ \\
$a_0/\Lambda^2 = +0.01 $ &   $77.2 \pm 0.1$ \\
$a_c/\Lambda^2 = -0.01 $ &   $75.5 \pm 0.1$  \\
$a_c/\Lambda^2 = +0.01 $ &   $76.9 \pm 0.1$ \\
$a_n/\Lambda^2 = -0.01 $ &   $76.0 \pm 0.1$  \\
$a_n/\Lambda^2 = +0.01 $ &   $76.0 \pm 0.1$  \\
\noalign{\smallskip}\hline
\end{tabular}
\end{table}

\begin{table}
\caption{The same as in Tab.~\ref{qagc200} at 
$\sqrt{s} = 500$~GeV.}
\label{qagc500}       
\begin{tabular}{llll}
\hline\noalign{\smallskip}
QAGC &  Cross section (fb) \\
\noalign{\smallskip}\hline\noalign{\smallskip}
Standard Model & $25.3 \pm 0.1$ \\
$a_0/\Lambda^2 = -0.001$ &  $83.8 \pm 0.3$ \\
$a_0/\Lambda^2 = +0.001$ &  $88.0 \pm 0.2$ \\
$a_c/\Lambda^2 = -0.001$ &  $41.3 \pm 0.2$ \\
$a_c/\Lambda^2 = +0.001$ &  $45.4 \pm 0.2$ \\
$a_n/\Lambda^2 = -0.001$ &  $26.4 \pm 0.1$ \\
$a_n/\Lambda^2 = +0.001$ &  $26.4 \pm 0.1$ \\
\noalign{\smallskip}\hline
\end{tabular}
\end{table}

\section{Conclusions}

The production of four fermions plus an additional detected photon
in $e^+e^-$ collisions is studied at LEP to test electroweak 
gauge boson couplings and in particular to derive bounds 
on QAGC. In order to provide predictions of phenomenological 
interest, an exact calculation of $4f + \gamma$ processes, 
including the effect of fermion masses, AGC and ISR has been 
performed. On the basis of the experimental accuracy, the 
contribution of fermion masses and ISR has been analyzed 
in comparison with typical deviations introduced by AGC.  
A new Monte-Carlo event generator ({\tt WRAP}) has been 
developed and is available for the simulation of 
radiative $4f$ events. 

The main conclusions of the present study can be summarized as follows. 
The effect of finite fermion masses, as analyzed in 
the $\mu^+ \nu_\mu \bar{c } s \gamma$ 
final state,    
turn out to be very sensitive to the separation angle 
$\vartheta_{\gamma-f}$ between photon and 
charged fermions, ranging from about 2\% for $\vartheta_{\gamma-f} = 5^\circ$ 
to about 7\% for $\vartheta_{\gamma-f} = 1^\circ$. For 
the realistic situation of a vanishing separation angle 
$\vartheta_{\gamma-\mu}$ a massive calculation is 
strictly unavoidable. 

Particular care has been devoted to the 
inclusion of ISR, as a consequence of the presence of an observed photon
in the final state. The contribution of ISR has been 
studied in terms of collinear and $p_\perp$ dependent SF. 
Numerical results illustrate that ISR  introduces corrections 
of the order of $10-15\%$ on the integrated cross section.
However, in order to get a reliable estimate of ISR corrections and 
to avoid double counting, 
$p_{\perp}$ photon effects have to be considered. It has been 
shown that the double counting, affecting the QED corrected cross section  
via collinear SF, may reach the 
$5\%$ level in a realistic event selection and   
hence it 
has to be taken into account carefully. 
A more accurate evaluation of double counting effects should however 
consider also the photonic radiation off final state charged fermions.

Both trilinear and genuinely quartic anomalous gauge couplings 
have been implemented in {\tt WRAP}, and their effects on total cross
section as well as on photon distributions have been investigated. 
The impact of TAGC on the considered observables does not seem to be very 
sensitive to the cuts imposed on the detected photon, suggesting that 
$W$ radiation is not easily disentangled from the fermion radiation. 
Thus $4f + \gamma$ final states are not the 
ideal place where to look for TAGC, if compared with $4f$ final states, 
which benefit of a higher statistics. On the contrary, these radiative 
processes are significantly affected by QAGC. 
In particular it has been shown that  
difference in the effect of QAGC are 
present between the predictions of the complete calculations by means 
of {\tt WRAP} and the ones obtained in the limit of on-shell 
$W$ bosons, which is the approximation presently used in the 
literature. 
A more complete investigation of QAGC in radiative events 
at $e^+ e^-$ colliders is currently in progress.

\noindent
{\bf Acknowledgements}  \par
The authors wish to thank 
F. Cavallari, D.G.~Charlton, A.~Denner, 
S.~Dittmaier, M.~Musy, C.G.~Papadopoulos, M.~Roth, D.~Wackeroth, 
and the participants in the CERN LEP2 Monte Carlo Workshop for useful 
discussions. A particular thank goes to the authors of {\tt RacoonWW} for 
the very helpful collaboration in numerical comparisons and to P.~Bell, 
D.G.~Charlton and M.~Thomson for useful information and help concerning 
QAGC.

\end{document}